\documentclass[aps, prd, twocolumn, lengthcheck, superscriptaddress, showpacs, letterpaper, nofootinbib]{revtex4-1}

\newcommand\sect[1]{\emph{#1.}---}

\usepackage{graphicx}
\usepackage{color}

\def\la{\langle}\def\ra{\rangle}
\def\be{\begin{eqnarray}}\def\ee{\end{eqnarray}}
\def\lsim{\mathrel{\rlap{\lower3pt\hbox{\hskip1pt$\sim$}}
     \raise1pt\hbox{$<$}}} 
\def\gsim{\mathrel{\rlap{\lower3pt\hbox{\hskip1pt$\sim$}}
     \raise1pt\hbox{$>$}}} 
\def\le{ \begin{array}{ll}}\def\re{\end{array}}

\def\lear{ \left( \begin{array}{cc}}\def\rear{\end{array} \right)}

\def\le{ \left( \begin{array}{cc}}\def\re{\end{array} \right)}

\def\bi{\bibitem}

\begin{document}

\title{Scale-Chiral Symmetry, Proton Mass  \\ and  Sound Velocity in Compact-Star Matter}

\author{Won-Gi Paeng}
\affiliation{%
Rare Isotope Science Project, Institute for Basic Science, Daejeon 305-811, Korea
}

\author{Mannque Rho}
\affiliation{%
Institut de Physique Th\'eorique, CEA Saclay, 91191 Gif-sur-Yvette c\'edex, France
}

\date{\today}

\begin{abstract}
With a light dilaton $\sigma$ and the light-quark vector mesons $V=(\rho,\omega)$ incorporated into an effective  scale-invariant hidden local symmetric Lagrangian, scale-chiral symmetry --  hidden in QCD -- arises at a high density, $n_{1/2}$,  as  an ``emergent" symmetry, a  phenomenon absent in standard chiral perturbative approaches but highly relevant for massive compact stars. What takes place  as the density increases beyond $n_{1/2}\sim 2n_0$ in compressed baryonic matter is (1)  a topology change from skyrmions to half-skyrmions, (2)  parity doubling  in the nucleon structure, (3) the maximum neutron star mass $M\simeq 2.05 M_{\odot}$ and the radius  $R\simeq 12.19$ km and (4)  the sound velocity $v_s^2/c^2\simeq 1/3$ due to  the ``vector manifestation (VM)" fixed point of $\rho$ and a ``walking" dilaton condensate, which is intricately connected  to the source of the proton mass.

\end{abstract}


\maketitle

\sect{Introduction}
We construct an effective field theory in which scale-invariant hidden local symmetry (HLS) is implemented and apply it to compressed baryonic matter. We show that with the ``bare" parameters of the effective Lagrangian matched to QCD at a suitable matching scale $\Lambda_M$, we can describe fairly reliably the bulk properties of nuclear matter and exploiting topological information obtained from skyrmion description of the Lagrangian -- that we shall refer to as $s$HLS, extrapolate the theory to high density appropriate for compact stars reaching densities $\sim (5-6)$ times the normal nuclear matter density $n_0\approx 0.16$ fm$^{-3}$. We find that for densities in the vicinity of that of the interior of  $\sim 2$-solar mass star, there is a strong hint for an emergent scale symmetry, which leads to  the sound velocity $v_s/c\approx 1/\sqrt{3}$ for massive compact stars and also accounts for {\it most} of the proton mass, with a mass-generation mechanism basically different from the standard paradigm anchored on spontaneously broken chiral symmetry.  Approaching, at high density, the dilaton-limit (DL)  fixed point  for the $\sigma$ and the vector manifestation (VM) fixed point for the $\rho$, scale-invariant hidden local symmetry manifests, giving rise to   ``mended symmetries" together with the massless $\pi$ and $a_1$.

\sect{Essential Ingredients and Assumptions}
Our strategy consists of combining what has been obtained in our previous works with new observations to arrive at new predictions. They are anchored on the following.
\begin{itemize}
\item (A) Our effective Lagrangian  combines scale symmetry with chiral symmetry,  putting a scalar Nambu-Goldstone, dilaton ($\sigma$), and  the psuedo-scalar pseudo-NG, pions ($\pi$) on the same putting. Our approach does not rely on a  precise scheme in implementing the dilaton  but we find the scheme proposed by Crewther and Tunstall~\cite{CT}  the most versatile for our purpose, as  explained in ~\cite{LPR-2016,LMR}. The resulting scale-chiral Lagrangian is then made hidden local symmetric by elevating the energy/mass scale to that of the lowest vector mesons $V=(\rho, \omega)$ following the argument of \cite{HY:PR,yamawaki}. We assume that the low-energy theorems are applicable in medium.
\item (B) The scale-chiral Lagrangian -- denoted $s$HLS in what follows -- is Wilsonian-matched  at the scale $\Lambda_M$ to QCD in terms of the well-known current correlators. This procedure renders the ``bare" parameters of $s$HLS dependent on the QCD condensates ${\cal C}= (\la\bar{q}q\ra, \la G^2\ra, ...)$ where $q$ and $G$ are, respectively, quark and gluon fields. The $s$HLS, when embedded in (nuclear) medium with density $n$, depends implicitly on density via its dependence on ${\cal C}$. We call this density dependence ``{\it intrinsic density (ID) dependence}." As defined, this ID dependence is Lorentz invariant.
\item (C) Baryons are brought into the theory  in two ways. One is to put the baryon fields explicitly into $s$HLS preserving scale-chiral symmetry. We call this Lagrangian $bs$HLS. Alternatively baryons can be generated as skyrmions from $s$HLS Lagrangian. This approach plays a key role for determining some of the important parameters of $bs$HLS that are unattainable otherwise.
\item (D) Up to nuclear matter density $n_0$, the available experimental data, accurately known up to lab momentum $\sim 300$ MeV, can be used to determine the ID dependence (IDD) of the bare parameters of $bs$HLS. This is feasible thanks to the low-energy theorems encoded in the Lagrangian in the vector, axial-vector and tensor channels.  The in-medium pion decay constant, $f_\pi^\ast$,  measured in deeply bound nuclear systems at $\sim n_0$~\cite{yamazaki} controls the density dependence of the parameters~\cite{dongetal,PKLR}. This procedure is equivalent to the standard chiral perturbative approach up to $n_0$~\cite{HRW}. However going above  $n_0$ is problematic because of the absence of guidance from both  experiments and theory.  We assume that the IDD so determined can be extrapolated,  at least,  slightly above $n_0$, say, $\sim 2n_0$. This density regime $0\lsim n\lsim 2n_0$ is referred to as Region-I (R-I for short).
\item (E)  The density regime  $n\gsim 2n_0$  is totally unknown, both experimentally and theoretically.  The most important IDD in the parameters of $bs$HLS  for $n \gsim 2n_0$, referred to as Region-II (R-II), can however be extracted by skyrmions simulated on crystal lattice.  It involves a topology change from skyrmions to half-skyrmions at a density denoted $n_{1/2}$. The existence of the half-skyrmion phase is predicted on the basis of symmetry~\cite{goldhaber-manton} and is actually seen in simulations on crystal~\cite{half-skyrmion-matter}. This prediction is robust and implementable in continuum treatments. Being topological, it involves only the pion field and is unaffected by other degrees of freedom. The half-skymion phase is characterized by that the quark condensate averaged over the unit cell, denoted $\Sigma$,  vanishes, although the condensate is non-zero locally, hence supporting chiral density wave. The pion decay constant is non-zero there, hence it is not in the Wigner-Weyl phase. The precise value for $n_{1/2}$ cannot be pinned down theoretically, but based on phenomenology, we estimate it to lie at $2\lsim n/n_0 \lsim 3$.
\item (F) With the $bs$HLS with the bare parameters with IDD determined in both regions, the energy density, $\epsilon (n)$, of the system is calculated using the $V_{lowk}$ renormalization-gorup (RG) technique as  employed in \cite{dongetal,PKLR}. The  EoS, the key element for compact-star physics, is gotten from $\epsilon$.
Now doing the relativistic mean-field calculation with $bs$HLS is equivalent to performing the double-decimation Landau Fermi-liquid calculation in the $V_{lowk}$ RG~\cite{BR:DD}.
\item (G) Treating nuclear matter in R-I is straightforward.  Both RMF of $bs$HLS and the $V_{lowk}$ derived from  $bs$HLS will do equally well.  They are equivalent in the sense first suggested by Matsui~\cite{matsui}. There is no need to resort to the skyrmion crystal model, which is not realistic,  in low-density regime. However in R-II as density increases, the half-skyrmion description becomes more realistic and predictive.  As described below, this allows one to map the results of the crystal simulation that capture mesonic and baryonic excitations in the {\it background} of the multi-soliton configuration to the IDD of the parameters of $bs$HLS.
\end{itemize}

\sect{Role of Topology Change}
The most important ingredient in our approach is the topology change from skyrmions to half-skyrmions in the crystal approach with $s$HLS at the density $n_{1/2}$. There are two outstanding observations  associated with this change. One is that it makes the nuclear tensor forces change drastically from below to above $n_{1/2}$. The other is that the matter above $n_{1/2}$ behaves like a Fermi liquid. Both seem to be totally at variance with standard nuclear models.

Consider the symmetry energy $E_{sym}$ that figures in the energy per particle of the system given as
\be
E(n,\alpha)=E_0 (n) +\alpha^2 E_{sym}(n) +O(\alpha^4)\label{Esym}
\ee  where $\alpha=(N-Z)/(N+Z)$ with $N(Z)$ standing for the neutron (proton) number. When calculated with skyrmions put on crystal lattice to simulate density effects, the $E_{sym}$ calculated  is found to have a cusp at $n_{1/2}$~\cite{LPR-cusp}. This cusp looks non-normal and may be taken as  an artifact of the crystal model. However it can be neatly explained by the behavior of the nuclear tensor forces at the cross-over density~\cite{LPR-cusp}. In fact this feature  gives a strong support for the role of topology in R-II as argued in \cite{LPR-cusp}.

In the $V_{lowk}$ approach (with $bs$HLS), the tensor forces -- which play the dominant role in $E_{sym}$ -- are given by the exchange of a pion and a $\rho$ with the parameters subject to IDD. Surprisingly they seem to be renormalization-group (RG) invariant, i.e.,  $(d/d\bar{\Lambda}) V_{lowk}^{tensor}=\beta([V_{lowk}^{tensor}], \bar{\Lambda})=0$ both in the vacuum and in medium~\cite{tensor-RGinvariant}. {This observation is made numerically, and remains to be proven rigorously. It resembles closely the flow equation for Landau Fermi-liquid fixed point interactions~\cite{shankar} although the tensor force is non-local unlike the Landau parameters which are local. }  If correct, then  the $E_{sym}$  could be accurately given by the fixed-point property both in R-I and in R-II.  We shall take this as an assumption to be confirmed. As observed in \cite{dongetal,PKLR}, the cusp in the skyrmion crystal description of $E_{sym}$ then suggests that the hidden gauge coupling $g_\rho$ goes over, at $n_{1/2}$,  from a constant to one that drops rapidly toward the vector manifestation (VM) fixed point $g_\rho\to 0$ in consistency with the vector manifestation of the $\rho$ meson at which $g_\rho\to 0$. This feature will be elaborated below. This changeover is responsible for $E_{sym}$ to go from soft to hard at $n_{1/2}$ that accounts for the massive stars~\cite{dongetal,PKLR}.

The second observation is the indication that in R-II, the half-skyrmion matter resembles Fermi-liquid. When the averaged quark condensate $\Sigma$ vanishes, there remains a large amount of non-vanishing mass, $m_{skyrmion}=m^\prime_0  + \Delta (\Sigma) \to m_0^\prime \neq 0$ as $\Sigma\to 0$ with the property~\cite{maetal-parity-doubling},
\be
m^\prime_0\propto f_\pi^\ast\propto \la\chi\ra^\ast \ \ \ {\rm with}\ \ \frac{\partial\, \la\chi\ra^\ast}{\partial n}=0 , \ \ n >  n_{1/2}\label{A}
\ee
which is reproduced in the mean field calculation of $bs$HLS\cite{paeng-interplay}.
This $m^\prime_0$ resembles the chiral-invariant nucleon mass $m_0$ in the parity-doublet model~\cite{detar-kunihiro}. In fact in heavy-light systems in medium, it is found that $m_0= m_0^\prime$~\cite{Ma-paritydouble}. The major difference is that here $m^\prime_0$ ``emerges" from strongly correlated system instead of put in {\it ab initio} by hand.

There are two points to stress here. First the part of the nucleon mass implied by $m_0^\prime\approx m_0$ is substantial, $\sim (0.7-0.9) m_N$  and points to the source of the proton mass, which is at variance with the standard paradigm for hadron mass based on spontaneously broken chiral symmetry~\cite{NJL}.  The standard paradigm  would suggest that the ratio of in-medium masses ${\cal R}=m_\rho^\ast/m_N^\ast\to {\rm constant} \sim 2/3$ as chiral restoration is approached. The prediction here is that ${\cal R}\to 0$.   Second the half-skyrmion crystal, when identified to be equivalent to the mean field of $bs$HLS, provides the IDDs of $bs$HLS Lagrangian in R-II through Eq.(\ref{A}).

It has remained mysterious since many years~\cite{BR:DD} that the hidden gauge coupling $g_\rho$ should be constant up to near $n_0$ and then drop rapidly to arrive at the VM fixed point (VMFP for short) $g_\rho=0$. This puzzle can be resolved  with the IDDs described above.

The nuclear matter at equilibrium can be described in terms of Wilsonian RG~\cite{shankar}. There the quasiparticle interactions have vanishing $\beta$ functions in the limit $N\equiv k_F/(\Lambda-k_F)\to \infty$ (where $\Lambda$ is the cutoff for decimation). With the vector mesons and the dilaton of $bs$HLS Lagrangian, considered heavy compared with the Fermi sea scale, integrated out to give the marginal four-point quasiparticle interactions, nuclear matter can be considered to be at its Fermi-liquid fixed point~\cite{song}.

Going beyond the equilibrium density, as argued, the Fermi-liquid structure should apply.   Now consider the parameter space of $bs$HLS on top of the Fermi-liquid fixed point. Approaching the IR fixed point with the scale parameter $\bar{\Lambda}\equiv \Lambda - k_F \rightarrow 0$, the parameters of the EFT Lagrangian should scale such that the $\beta$ function for the quasiparticle interactions be zero  at a given Fermi-momentum $k_F$. Suppose the density is changed from $k_{F1}$ to $k_{F2}$. Then certain parameters should change, say the quasiparticle mass as an example, from $m^\ast(k_{F1},\,\bar{\Lambda}=0)$ to $m^\ast(k_{F2},\,\bar{\Lambda}=0)$ to preserve $\beta(k_{F1},\, \bar{\Lambda}=0) = \beta(k_{F2},\, \bar{\Lambda}=0) = 0$.  This means that the Fermi-liquid fixed point quantities are closely related to each other at  given density so that $g_V^\ast(k_F,\,\bar{\Lambda}=0)$ as well as $m^\ast(k_F,\,\bar{\Lambda}=0)$ should be dependent on $\langle \chi \rangle^\ast$ and $k_F$ to have $\beta(k_F,\,\bar{\Lambda}=0)=0$.
Thus in the density regime $n \lsim n_{1/2}$ (R-I), the condensate $\langle \chi \rangle^\ast$ given in the mean field~\cite{paeng-interplay}, locked to the quark condensate  $\langle\bar{q}q\rangle$, decreases as observed in experiments~\cite{yamazaki}.

Now going to $n > n_{1/2}$,  the dilaton condensate $\langle \chi \rangle^\ast$ should stay constant as predicted by the theory, i.e., (\ref{A}). This requires that $g_{\rho,\omega}^\ast(k_F,\,\bar{\Lambda}=0)$ scale to preserve  $\beta(k_F,\, \bar{\Lambda}=0)=0$ as density increases.
This interplay between the dilaton condensate and the vector coupling for the Fermi-liquid fixed point structure turns out to bring about  the change in density dependence of $g_\rho^\ast$ from $g_\rho^\ast \approx g_\rho$ near $n=n_0$ to $g_\rho^\ast \rightarrow 0$ near the VM fixed point.

\sect{Dilaton-Limit Fixed Point (DLFP)}%
Given a Lagrangian of the form of $bs$HLS, one can gain further insight into the structure of dense matter in terms of scale-chiral symmetry without resorting to the input from the topology change discussed above. In ~\cite{Paeng:2011hy,paeng-interplay}, it was shown in terms of Wilsonian renormalization group (RG) equations that there is a possible infrared fixed point in $bs$HLS
\begin{equation}
\left(m_N,\, g_{V\rho} -1,\, g_{A} -g_{V\rho} \right) \rightarrow \left( 0,\, 0,\, 0 \right) \label{DLFP}
\end{equation}
for the nucleon mass $m_N$, the $\rho NN$ coupling $g_{\rho NN} = \left(g_{V\rho} -1\right)g_\rho$ -- with $g_\rho$ the hidden gauge coupling and $g_{V\rho}-1$ a medium-renormaization -- and $g_A$ for the axial-vector current.
 The parameters of the Lagrangian satisfy a set of coupled renormalization-group equations and can flow in various different directions.  We consider how the fixed point (\ref{DLFP}) can be reached in dense medium.

In the mean field,  equivalent to Landau fixed-point theory,    there are singular terms when {$\langle \chi \rangle^\ast$ goes to zero
arising from the non-linear terms of $N$, $\bar{\sigma} \equiv \langle \chi \rangle^\ast \sigma/f_\sigma $ and $\bar{\pi} \equiv \langle \chi \rangle^\ast \pi/f_\sigma$.  Taking the limit $\langle \chi \rangle^\ast \rightarrow 0$ corresponds to the DLFP, where $f_\pi = f_\sigma$ -- which also follows from the locking of two symmetries in the CT theory~\cite{CT} --  and $g_A = g_{V\rho}$ are required} to avoid the singularity.  In this limit, one obtains a Gell-Mann-L\'evy (GML)-€type linear sigma model with the degenerate $O(4)$ multiplet for the scalar and pseudo-scalar fields (massless in the chiral limit)~\cite{paeng-interplay,Beane:1994ds,Sasaki:2011ff}. There  the $\rho$ mesons are decoupled from the nucleons.

\sect{``Walking" Dilaton Condensate}%
The dilaton condensate in the half-sykrmion phase with the property of (\ref{A}) can be reproduced in the mean-field of $bs$HLS in terms of an interplay between the $\omega$ mass and the dilaton condensate,  the former blocking the  dropping of the latter $\propto f_\pi^\ast$ in R-I, as one goes beyond $n_{1/2}$. This is illustrated in \cite{paeng-interplay}.

Now the question is: How is the dilaton limit fixed point approached with the ``walking" dilaton condensate ``running" to zero? The answer is in the many-body correlations inducing $(g_{V\rho}-1, g_{V\omega} -1)\to (0, 0)$ as the DLFP is approached.  It is not in the IDDs in contrast to the VM fixed point which involves the hidden gauge coupling intrinsic to QCD.   As the dilaton condensate goes to zero toward the DLFP, we have the scale symmetry restored, $\theta_\mu^\mu=\partial_\mu D^\mu\to 0$, but it is an ``induced" symmetry restoration. This resembles the scenario of chiral symmetry restoration involving a soft scalar mode~\cite{kunihiro}.

\sect{Mended Symmetries}
The DLFP brings $\pi$ and $\sigma$ together while the VMFP brings $\pi$ and $\rho$ together. The two fixed points could come together or separately. If they come separately, the former should come before the latter since DLFP is driven by correlated interactions without the gauge coupling going to zero. Furthermore it has been shown that the axial vector meson $a_1$ could also come together with the $\rho$ at the VMFP~\cite{genHLS}. This means that $\pi$, $\sigma$, $\rho$ and $a_1$ could all come together at some density which may or may not coincide with chiral restoraion. In the chiral limit, this would involve massless gauge bosons -- $\rho$ and $a_1$ -- together the Nambu-Goldstone bosons $\pi$ and $\sigma$. An attractive possibility is that they form  the multiplet of  Weinberg's ``mended symmetries"~\cite{mended}.

\sect{Predictions on Compact Stars}
We now apply the $bs$HLS with the IDDs fixed as above to the properties of compact stars.

Consider the in-medium vacuum expectation value of the energy-momentum tensor $\langle \theta^\mu_{\,\mu} \rangle = \epsilon(n) -3 P(n)$ expressed in terms of the thermodynamic qunatities, the energy density $\epsilon(n)$ and  the pressure $P(n)$.   The energy-momentum tensor and the sound velocity are related by
$\frac{\partial}{\partial n} \langle \theta^\nu_{\,\nu} \rangle = \frac{\partial\, \epsilon(n)}{\partial n}\left(1 -3 \frac{v_s^2}{c^2}\right)$
where we used $\frac{v_s^2}{c^2}=\frac{\partial\,P(n)}{\partial n} /\frac{\partial\,\epsilon(n)}{\partial n}$ for the sound velocity $v_s$.
Now if $\langle \theta^\mu_{\,\mu} \rangle$ is a constant independent  of density, given that  the energy-density of compact-star matter has no known extremum, i.e., $\frac{\partial\, \epsilon(n)}{\partial n}\neq 0$,  then it will follow that $v_s/c = \sqrt{1/3}$.

Let us first look at the mean-field approximation in R-II. The trace of the energy-momentum tensor in dense matter in the mean-field approximation with the $bs$HLS Lagrangian  is given by {$\langle \theta^\mu_\mu \rangle \propto \langle\chi\rangle^{\ast\,4} $}. This is because to the leading order in the scale-chiral counting, $O(p^2)$, the energy density is scale-invariant since $bs$HLS is scale-invariant, the scale symmetry breaking being only in the dilaton potential. Therefore if $\langle \chi \rangle^\ast$ stays constant in $n\gsim n_{1/2} $ as seen in (\ref{A}), the sound velocity will become $v_s/c = \sqrt{1/3}$.

 Now what happens if one goes beyond the mean-field approximation? Given that the mean-field approximation is presumably equivalent to doing Fermi-liquid fixed theory, going beyond the mean-field must then be tantamount to calculating higher loop corrections to the Landau fixed point theory. Such a calculation using the $V_{lowk}$ RG  double decimation has been formulated and applied to compact-star matter in \cite{PKLR}. We have applied the same formalism with only a slight modification of the approach to the VM fixed point -- which is of course unknown and hence a free parameter --  in \cite{PKLR} (without affecting the known properties of nuclear matter well described therein ).

\begin{figure}[h]
\begin{center}
\includegraphics[height=3.8cm]{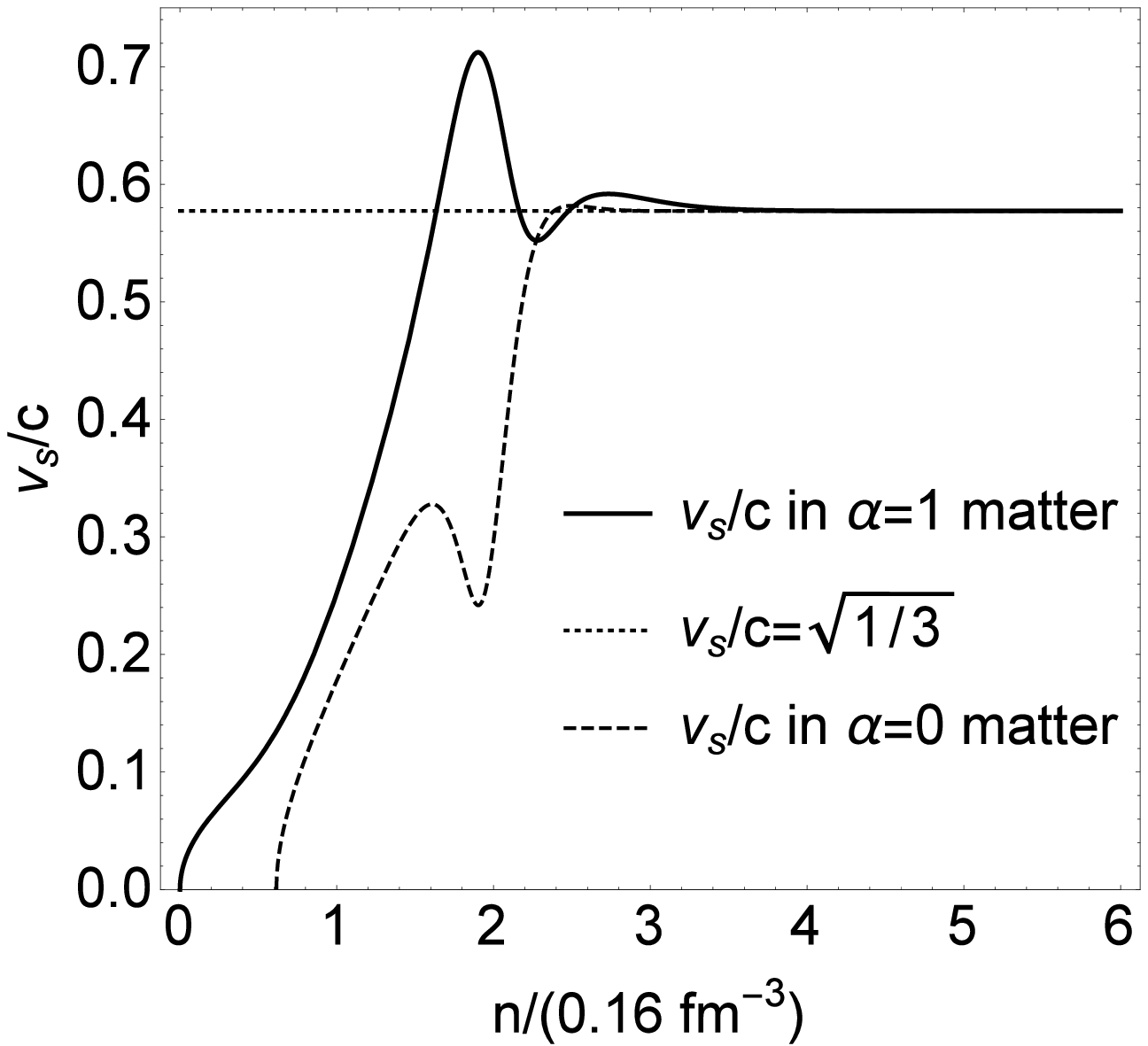}\includegraphics[height=3.8cm]{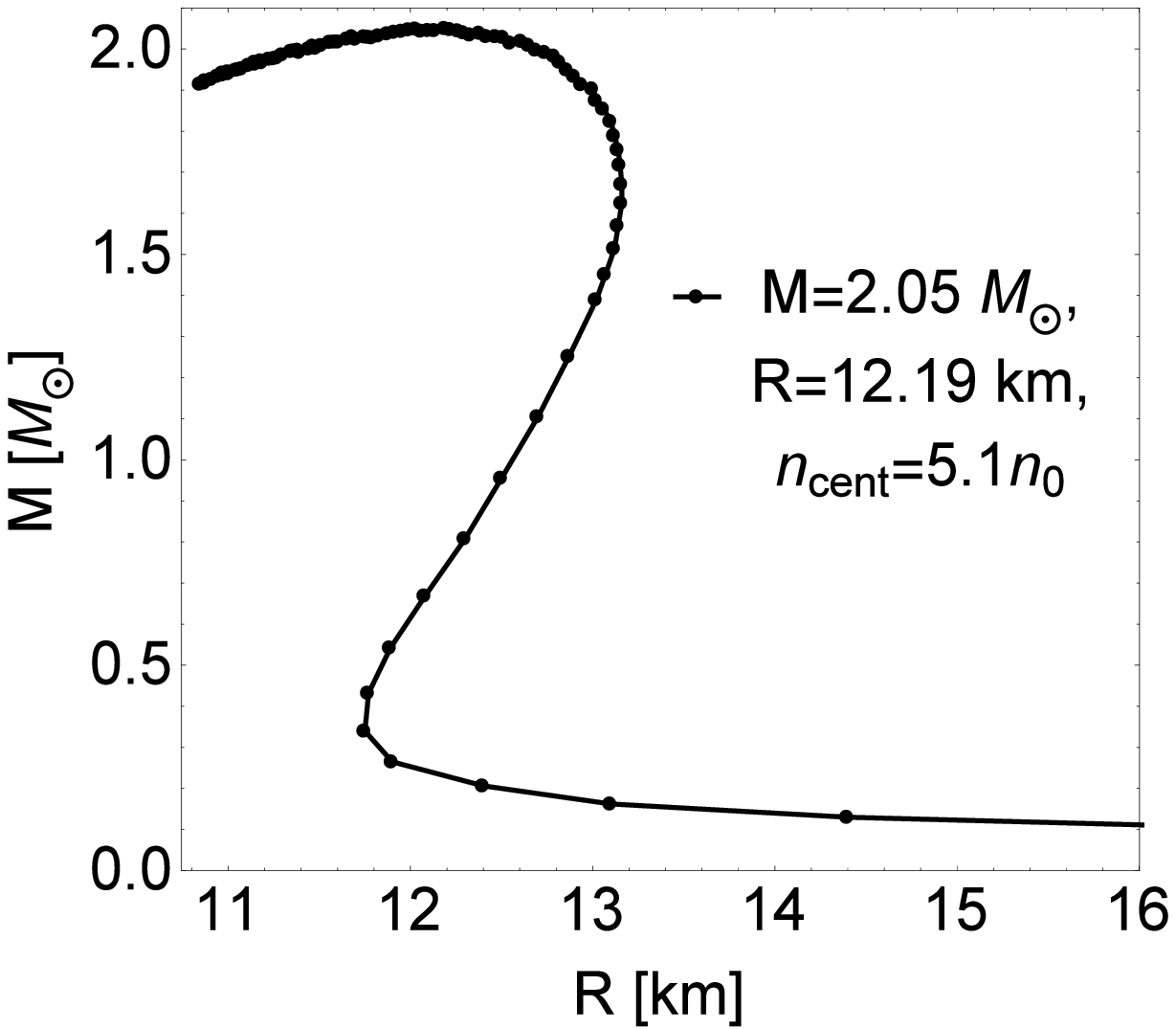}
\caption{
Left panel: The sound velocity predicted in the $V_{lowk}RG$ with $bs$HLS for $\alpha=\frac{N-Z}{N+Z}$. The oscillation of $v_s$ near $n_{1/2}$ is considered to be an artifact of the sharp changeover of the states before and after $n_{1/2}$.  Right panel:  (M)ass vs. (R)adius of the neutron star with the beta equilibrium taken into account.}
\label{prediction}
\end{center}
\end{figure}

A surprisingly simple way of parameterizing the EoS given by the $V_{lowk}$ RG procedure is to express the energy per particle of the dense matter in R-II by
\be
E/A=-m_N + B\Big(\frac{n}{n_0}\Big)^{1/3} + D\Big(\frac{n}{n_0}\Big)^{-1}\label{parameter}
\ee
where $m_N$ is the nucleon mass, $B$ and $D$ are constants of mass dimension. One can show that (\ref{parameter}) is the solution of $\frac{d\,P}{dn} = \frac{1}{3} \frac{d\,\epsilon}{dn}$ which assumes the constant $\langle \theta^\mu_\mu\rangle$.  The strategy then is to pick the coefficients $B$ and $D$ for $\alpha=(0,1)$  to reproduce the $E_0$ and $E_{sym}$ in (\ref{Esym}) that reproduces accurately those obtained from the $V_{lowk}$RG with $bs$HLS with the IDDs that encode the ``walking condensate" associated with the DLFP  and VMFP in R-II. The induced effects that accompany the DLFP, namely in $g_{V\rho}$ and $g_{V\omega}$, are captured in the $V_{lowk}$ approach in the RG decimations, extraneous to IDD. This strategy can be made to work extremely well. Indeed very accurate fits can be  obtained for both $E_0$ and $E_{sym}$ in the density range $2\lsim n/n_0\lsim 6$ relevant for massive compact stars.\footnote{ The fit parameters are $B_{\alpha=(0,1)}= (570.031 \textnormal{ MeV},\, 686.4 \textnormal{ MeV})$ and $D_{\alpha=(0,1)}= (439.676 \textnormal{ MeV},\, 252.888\textnormal{ MeV})$. For lack of space the detailed fits will be shown elsewhere.} This means that the star matter can have $v_s/c\simeq  1/\sqrt{3}$ for density regime $n > n_{1/2}$.  Given in Fig.~\ref{prediction} are the results {\it predicted} in the $V_{lowk}$RG with $bs$HLS for the sound velocity for both neutron and symmetric nuclear matter and $M$ vs. $R$.  The result for $v_s$ is totally unprecedented and intriguing: there is no other (known to us) trustful theory that gives {$v_s/c < 1/\sqrt{2}$}!
Given that there is still quite a bit of room for adjustments in the IDDs, quantitative comparison with observed data may not be very meaningful.  Nonetheless  the results obtained in this approach are satisfactory and encouraging: Maximum star mass $M\simeq  2.05 M_\odot$, radius $R\simeq 12.19$ km, central density $n_{cent}\simeq  5.1n_0$, $E_{sym} \simeq 26$ MeV and $L\simeq  49$ MeV at equilibrium density. For completeness, we also list the predictions for normal nuclear matter which, as mentioned, should be comparable to those of the standard chiral perturbation approach: Equilibrium density $n_{\rm eq}\simeq  0.15n_0$, B.E. $\simeq 16$ MeV, compression modulus $K\simeq 215$ MeV.

Details will be reported elsewhere.

\vskip 0.2cm
We are grateful to Tom Kuo and Hyun Kyu Lee for valuable discussions and to Kyungmin Kim for the TOV code used in the star calculation. The work of WGP is supported by the Rare Isotope Science Project of Institute for Basic Science funded by Ministry of Science, ICT and Future Planning and National Research Foundation of Korea (2013M7A1A1075764).


\end{document}